\documentclass[twocolumn]{aastex631}

\usepackage{amsmath}
\usepackage{graphicx}
\usepackage{float}

\shorttitle{Characterization of TOI 7475.01}
\shortauthors{Escolà-Rodrigo}

\begin{document}

\title{Statistical Validation and Photometric Characterization of the Hot Jupiter Candidate TOI 7475.01}

\author[0009-0007-1479-2386]{Biel Escolà-Rodrigo}
\affiliation{Independent Researcher}

\begin{abstract}
We present the statistical validation and full photometric characterization of the exoplanet candidate TOI 7475.01 (TIC 376866659), detected by the TESS mission. Using a custom pipeline combining natural flux preservation with robust BLS detection, we identified a transit signal with a period of $3.2538$ days and a depth of $\sim 4600$ ppm. To rule out false positives, we performed centroid analysis, spatial contamination checks using Gaia DR3, and a statistical validation using \texttt{TRICERATOPS}. Our results show a Signal-to-Noise Ratio (SNR) of 294.13 and a False Positive Probability (FPP) of $\approx 0$. Based on the clean spatial environment, stable centroids, and high statistical probability, we validate TOI 7475.01 as a planetary companion. We subsequently performed a Bayesian transit fit using \texttt{juliet} with Dynesty nested sampling, and derived physical parameters via Monte Carlo error propagation. We find a planetary radius of $R_p = 1.18^{+0.39}_{-0.40}\ R_\mathrm{Jup}$ and an equilibrium temperature of $T_\mathrm{eq} = 1455^{+77}_{-56}$ K, consistent with a hot Jupiter classification. The planet mass is estimated at $M_p \approx 2.2\ M_\mathrm{Jup}$ (MAP) via the \citet{chen2017} mass-radius relation; radial velocity follow-up is required for a precise measurement. The impact parameter remains unconstrained ($b = 0.46 \pm 0.34$), a limitation of single-band TESS photometry that future CHEOPS observations could resolve.
\end{abstract}


\section{Introduction}

The Transiting Exoplanet Survey Satellite (TESS) has revolutionized the search for exoplanets around nearby bright stars \citep{ricker2015}. In this work, we focus on TOI 7475.01, a candidate listed on the Exoplanet Follow-up Observing Program (ExoFOP) website \citep{exofop}, associated with TIC 376866659.

This paper combines two stages of analysis: the statistical validation of the candidate (Sections~\ref{sec:detection}--\ref{sec:triceratops}), which confirms the planetary nature of the signal, and a full photometric characterization (Sections~\ref{sec:fit}--\ref{sec:discussion}), which derives the physical properties of the planet via Bayesian modeling. The goal of the first stage is to vet the signal against common astrophysical false positives, such as eclipsing binaries (EBs) and background contamination. The goal of the second stage is to determine the orbital and physical parameters of the validated companion.

\section{Detection and Light Curve Analysis}
\label{sec:detection}

We processed the TESS Sector 91 data using the \texttt{Lightkurve} package \citep{lightkurve2018} with a custom pipeline that avoids aggressive flattening to preserve transit geometry. We applied a Box-Least Squares (BLS) algorithm \citep{kovacs2002} with a multi-duration search grid.

The detection yielded a clear signal with a Signal-to-Noise Ratio (SNR) of \textbf{294.13}. The phase-folded light curve exhibits a distinct \textbf{U-shaped transit} with a flat bottom, which is characteristic of a planetary body transit rather than the V-shape typical of grazing eclipsing binaries.

\begin{figure}[ht!]
    \centering
    \includegraphics[width=1.0\linewidth]{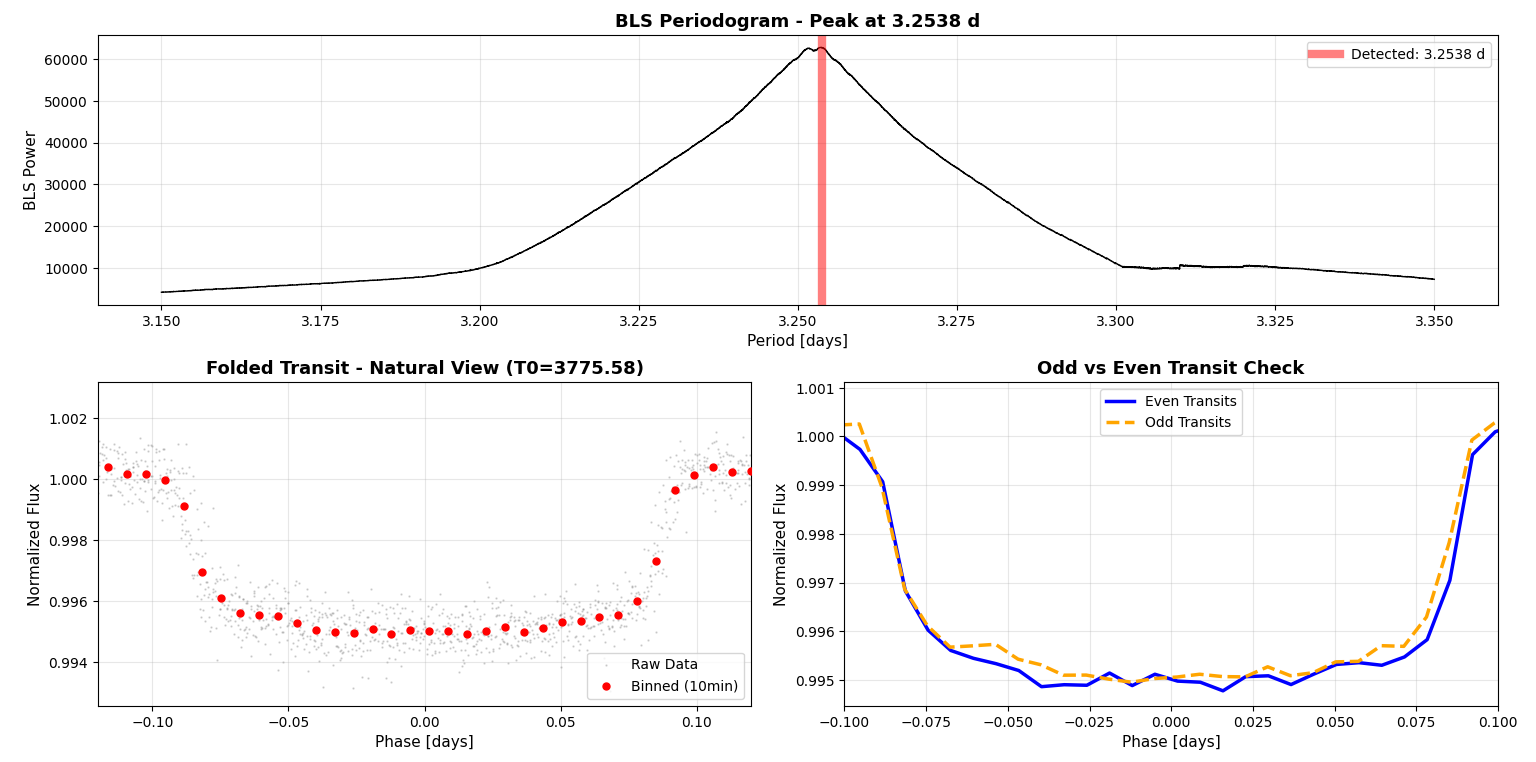} 
    \caption{\textbf{Detection Summary.} Top: BLS Periodogram showing a strong peak at 3.25 days. Bottom Left: Phase-folded light curve (red points represent 10-minute bins) showing a clear U-shape. Bottom Right: Odd/Even transit check, confirming consistent depths and ruling out secondary eclipses.}
    \label{fig:detection}
\end{figure}

Furthermore, the Odd/Even transit check (Figure \ref{fig:detection}, bottom right) confirms that consecutive transits have consistent depths. This lack of depth alternation strongly disfavors a binary star scenario. The derived parameters are summarized in Table \ref{tab:params}.

\begin{table}[h!]
    \centering
    \caption{Derived Planetary Parameters (Sector 91)}
    \begin{tabular}{lc}
        \hline \hline
        Parameter & Value \\
        \hline
        Period ($P$) & 3.253773 days \\
        Epoch ($T_0$) & 3775.5819 BTJD \\
        Transit Depth & 4601 ppm \\
        Duration & 4.08 hours \\
        SNR & 294.13 \\
        Noise Level & 477 ppm \\
        \hline
    \end{tabular}
    \label{tab:params}
\end{table}

\section{Vetting and False Positive Exclusion}
\label{sec:vetting}

To assess the nature of the candidate, we employed a multi-stage vetting protocol focusing on spatial contamination and centroid motion.

\subsection{Spatial Contamination (Gaia DR3)}
We analyzed the stellar neighborhood using Gaia DR3 data \citep{gaia2023} to check for potential background eclipsing binaries that could be contaminating the photometric aperture.

\begin{figure}[ht!]
    \centering
    \includegraphics[width=0.8\linewidth]{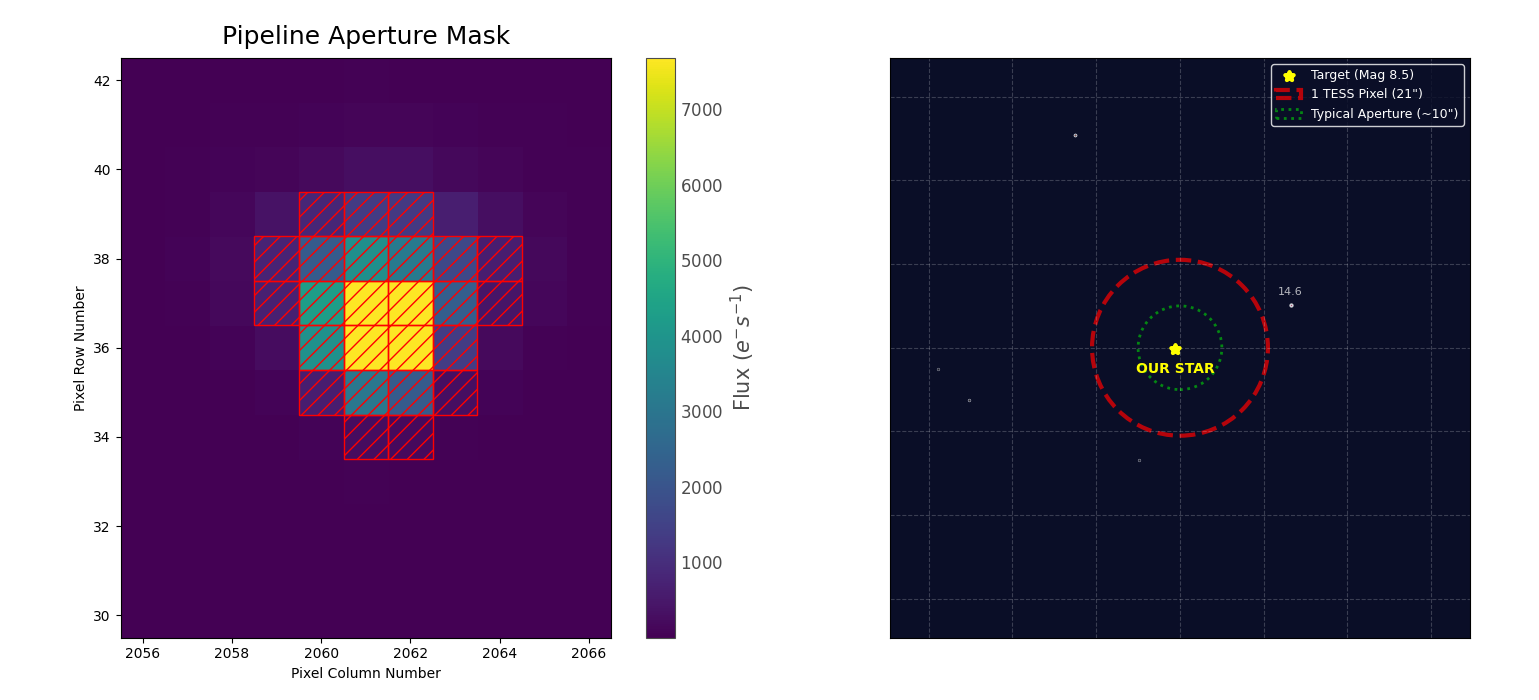}
    \caption{\textbf{Gaia DR3 Star Map.} The target (Gold Star) is centered. The red dashed circle represents the size of one TESS pixel (21"). The nearest neighbor (white dot) is 28.3" away, well outside the critical area, confirming a clean aperture.}
    \label{fig:starmap}
\end{figure}

As shown in Figure \ref{fig:starmap}, the field is remarkably clean. The target star (TIC 376866659) is isolated within the typical TESS pixel scale ($\sim 21$ arcseconds).
\begin{itemize}
    \item \textbf{Nearest Neighbor:} The closest detected source is located at a separation of \textbf{28.3 arcseconds}, which places it outside the critical contamination zone of the central pixel.
    \item \textbf{Magnitude Contrast:} This neighbor is significantly fainter ($G = 14.56$) compared to the target ($G = 8.48$), resulting in a contrast of $\Delta G \approx 6$. It is too faint to mimic the observed transit depth.
    \item \textbf{Target Stability (RUWE):} The target star has a Renormalized Unit Weight Error (RUWE) of \textbf{1.02}. A value close to 1.0 indicates a good astrometric fit, suggesting the target is a single star and not an unresolved close binary.
\end{itemize}

We conclude that the signal is not caused by spatial contamination from a known background source.

\subsection{Centroid Analysis}
We performed a centroid shift test using the moments method provided by \texttt{Lightkurve} to verify if the photocenter of the target moves during the transit event. A significant shift during the eclipse would indicate that the flux dip is caused by a background object rather than the target star.

\begin{figure}[ht!]
    \centering
    \includegraphics[width=1.0\linewidth]{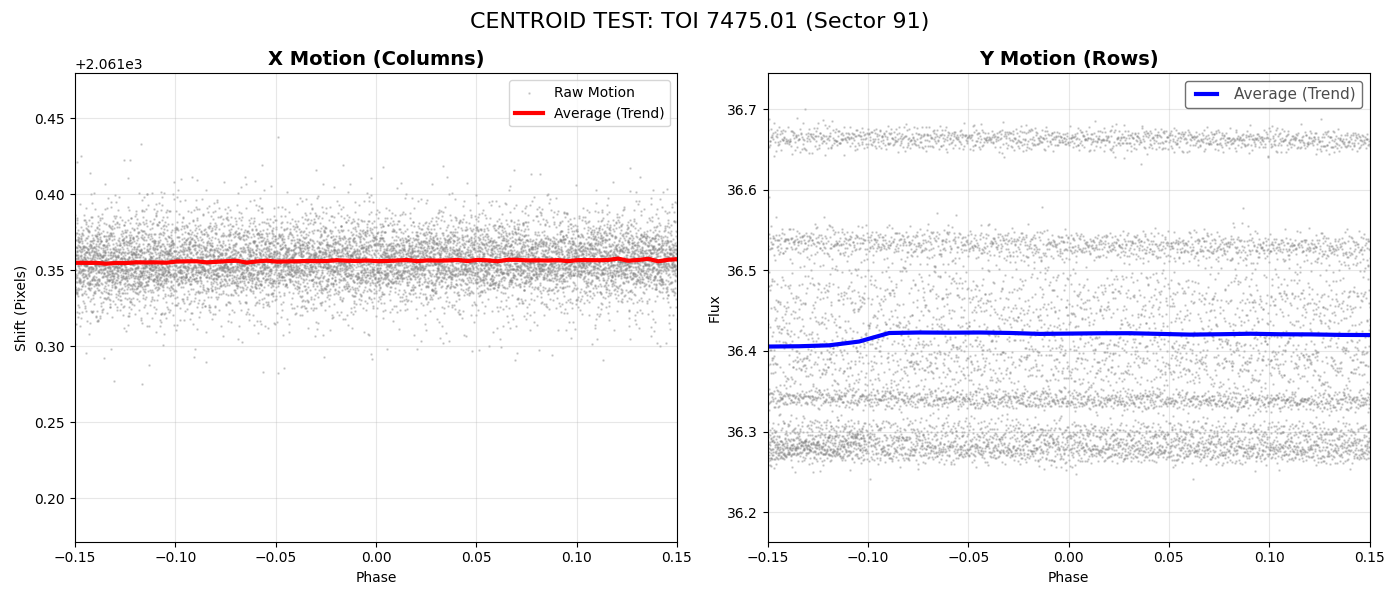}
    \caption{\textbf{Centroid Motion Analysis.} Left: X/Column movement. Right: Y/Row movement. The colored lines (red/blue) represent the binned average trend. The lines are essentially flat during the transit, ruling out background binaries.}
    \label{fig:centroids}
\end{figure}

We processed 95,646 valid data points from the Target Pixel File (TPF) and calculated the center of light. Figure \ref{fig:centroids} shows the X (column) and Y (row) position of the centroid phase-folded to the planetary period.

\begin{itemize}
    \item \textbf{Result:} The average centroid trend (solid lines) remains flat and stable during the transit window (Phase 0.0).
    \item \textbf{Interpretation:} While a minor fluctuation is observed at the beginning of the Y-axis plot, it is attributable to instrumental jitter and does not mirror the distinct U-shape of the transit flux. 
\end{itemize}

The absence of a correlated shift confirms that the transit signal originates from the target star, TIC 376866659.

\section{Statistical Validation (TRICERATOPS)}
\label{sec:triceratops}

To statistically validate the planetary nature of the candidate, we utilized the \texttt{TRICERATOPS} package \citep{giacalone2021}. This tool calculates the False Positive Probability (FPP) by simulating various astrophysical scenarios based on the TESS light curve, aperture geometry, and Gaia DR3 stellar parameters.

Given the stochastic nature of the MCMC analysis, we performed a robustness test consisting of \textbf{20 independent runs}. This ensures that the derived probabilities are stable.

\subsection{Results}
The analysis included the target star (TIC 376866659) and 28 surrounding sources within a 2.5 arcminute radius. The results across all 20 runs were remarkably consistent, yielding a mean False Positive Probability (FPP) effectively of zero.

\begin{table}[h!]
    \centering
    \caption{Triceratops Validation Statistics (20 Runs)}
    \begin{tabular}{lccc}
        \hline \hline
        Metric & Mean & Std Dev & Threshold \\
        \hline
        \textbf{FPP} (False Positive) & \textbf{0.000000} & $\pm 10^{-6}$ & $< 0.015$ \\
        \textbf{NFPP} (Nearby FPP) & 0.000000 & $\pm 0.0$ & $< 0.001$ \\
        \hline
    \end{tabular}
    \label{tab:triceratops_stats}
\end{table}

As shown in Table \ref{tab:triceratops_stats}, the FPP is orders of magnitude below the standard validation threshold of $1.5\%$ ($0.015$). This statistically rules out astrophysical false positives such as Eclipsing Binaries (EB) or Background Eclipsing Binaries (BEB).

\subsection{Scenario Probability Breakdown}
Since the probability of the signal being a False Positive is negligible, the remaining probability mass describes the likelihood of the signal being a planet in different configurations. The distribution of the winning scenarios across the 20 runs is summarized as follows:

\begin{itemize}
    \item \textbf{TP: $\sim 45\%$.} The planet orbits the target star. This is the most likely single scenario.
    \item \textbf{PTP: $\sim 40\%$.} The planet orbits an unresolved bound companion. In this scenario, the planet still belongs to the target system.
    \item \textbf{DTP: $\sim 15\%$.} An unresolved background star is present in the aperture, but the transiting planet orbits the \textbf{target star}. Although the presence of an unresolved background star seems unlikely given the clean Gaia DR3 data (Section~\ref{sec:vetting}), even in this scenario, the transit occurs on the target.
\end{itemize}

\begin{figure}[ht!]
    \centering
    \includegraphics[width=1.0\linewidth]{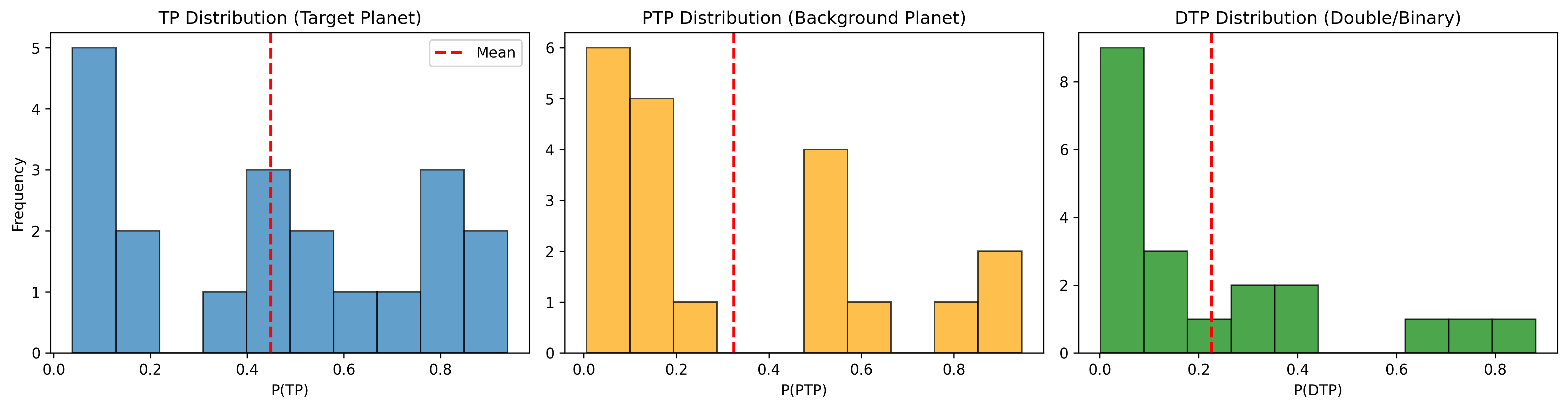}
    \caption{Probability distribution of the three dominant scenarios over 20 runs. The TP (Target Planet) and PTP (Bound Companion Planet) scenarios dominate. Note that for DTP, the transit also occurs on the target star, confirming the signal's origin.}
    \label{fig:triceratops_hist}
\end{figure}

Crucially, in all three dominant scenarios (TP, PTP, and DTP), the planet is orbiting the target star or a bound companion. The cumulative probability that the planet belongs to the TIC 376866659 system is therefore $\approx 100\%$ (Figure \ref{fig:triceratops_hist}).

\section{Bayesian Transit Fit}
\label{sec:fit}

\subsection{Light Curve Preprocessing}

\begin{figure}[ht!]
    \centering
    \includegraphics[width=1.0\linewidth]{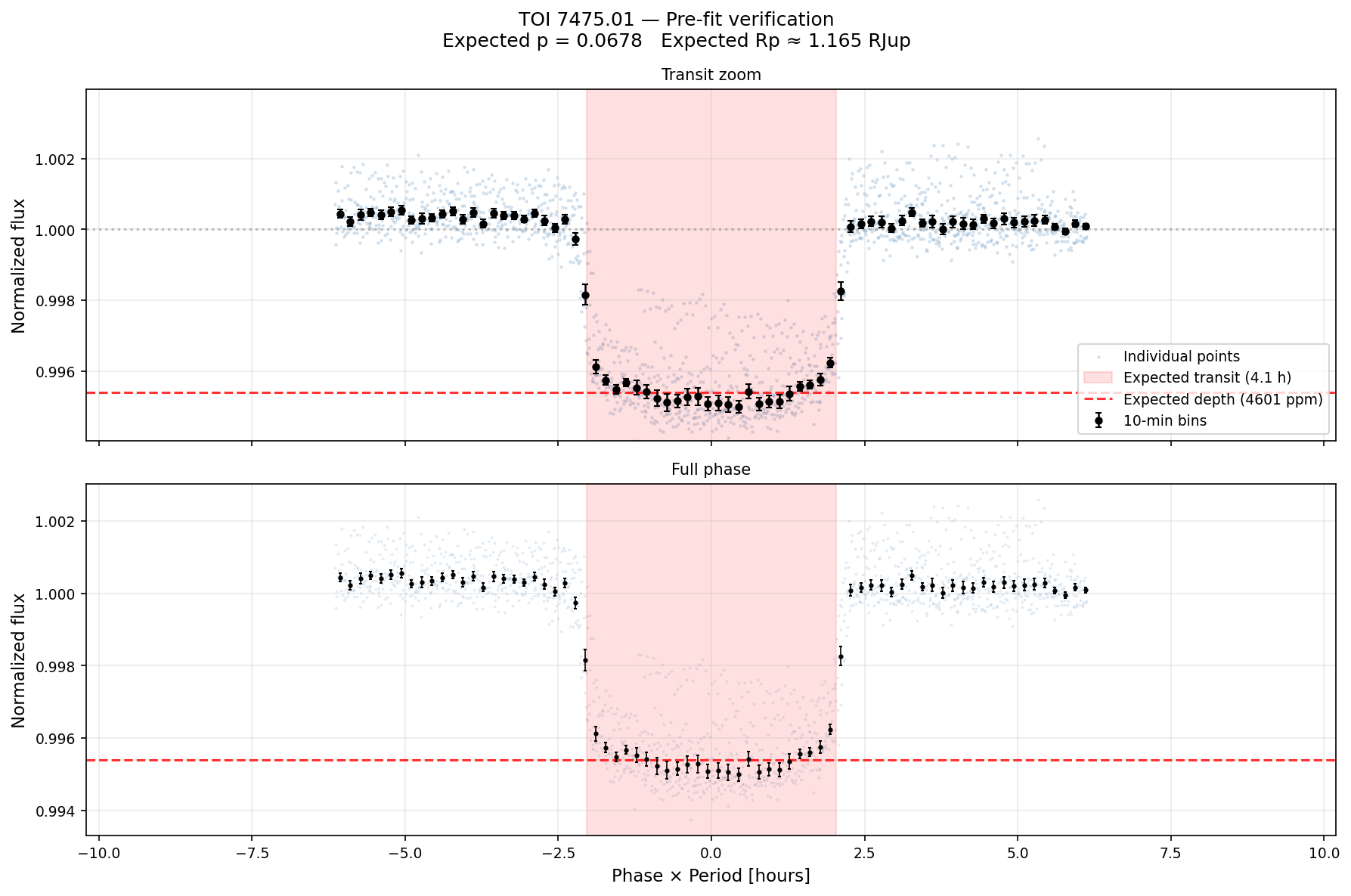}
    \caption{\textbf{Pre-fit Phase-folded Light Curve.} Top: zoom around the transit. Black points are 10-minute bins; the red shaded region marks the expected transit window; the dashed line marks the expected depth. Bottom: full phase view showing the out-of-transit baseline. The transit is clearly detected at the expected depth and epoch.}
    \label{fig:phased}
\end{figure}

For the Bayesian fit we applied transit windowing, retaining only data within $1.5 \times T_{14}$ of each predicted transit center, reducing the dataset to 1,641 binned data points from the original 86,906-point SPOC light curve. The pre-fit flux diagnostic confirms robust transit detection: the observed in-transit flux decrement is $-4735$ ppm, consistent with the BLS depth of $-4601$ ppm (3\% discrepancy, within noise). Figure~\ref{fig:phased} shows the phase-folded light curve with 10-minute binning.

\subsection{Model and Priors}

We performed a Bayesian transit fit using \texttt{juliet} \citep{espinoza2019}, which interfaces with the Dynesty nested sampler \citep{speagle2020}. We adopted a quadratic limb darkening law in the \citet{kipping2013} parametrization, with Gaussian priors on the coefficients centered on the \citet{claret2017} theoretical values for the stellar parameters of TIC 376866659: $q_1 = 0.348 \pm 0.05$, $q_2 = 0.246 \pm 0.05$.

The eccentricity was fixed to zero, justified by the short orbital period ($P = 3.25$ days) at which tidal circularization timescales are expected to be shorter than the system age \citep{goldreich1966}.

The stellar mean density $\rho_* = 350.9 \pm 93.1$ kg m$^{-3}$ (computed from $M_*$ and $R_*$) was included as a free parameter with a Gaussian prior. This allows \texttt{juliet} to compute $a/R_*$ self-consistently via Kepler's third law reformulated in terms of density:
\begin{equation}
    \frac{a}{R_*} = \left(\frac{G\,\rho_*\,P^2}{3\pi}\right)^{1/3},
\end{equation}
which links the transit light curve directly to a stellar observable independent of the planet.

The prior on the radius ratio $p = R_p/R_*$ was set uniform over $[0.034, 0.102]$, bracketing $0.5\times$ to $1.5\times$ the value expected from the BLS depth ($p_\mathrm{exp} = \sqrt{\delta} = 0.068$). All prior settings are summarized in Table~\ref{tab:priors}.

The nested sampling was run with 1,000 live points and convergence criterion $\Delta\log Z < 0.1$, requiring approximately 323,000 likelihood evaluations.

\begin{table}[h!]
    \centering
    \caption{Prior Distributions for the Bayesian Fit}
    \begin{tabular}{lll}
        \hline \hline
        Parameter & Prior & Value \\
        \hline
        $P$ [days]       & Normal    & $3.253773 \pm 0.0001$ \\
        $T_0$ [BTJD]     & Normal    & $3775.582 \pm 0.005$ \\
        $p = R_p/R_*$    & Uniform   & $[0.034,\ 0.102]$ \\
        $b$              & Uniform   & $[0.0,\ 0.99]$ \\
        $\rho_*$ [kg/m³] & Normal    & $350.9 \pm 93.1$ \\
        $e$              & Fixed     & $0$ \\
        $q_1$            & Normal    & $0.348 \pm 0.05$ \\
        $q_2$            & Normal    & $0.246 \pm 0.05$ \\
        $\sigma_w$ [ppm] & Log-unif. & $[0.1,\ 1000]$ \\
        \hline
    \end{tabular}
    \label{tab:priors}
\end{table}

\subsection{Posterior Constraints}

\begin{figure}[ht!]
    \centering
    \includegraphics[width=1.0\linewidth]{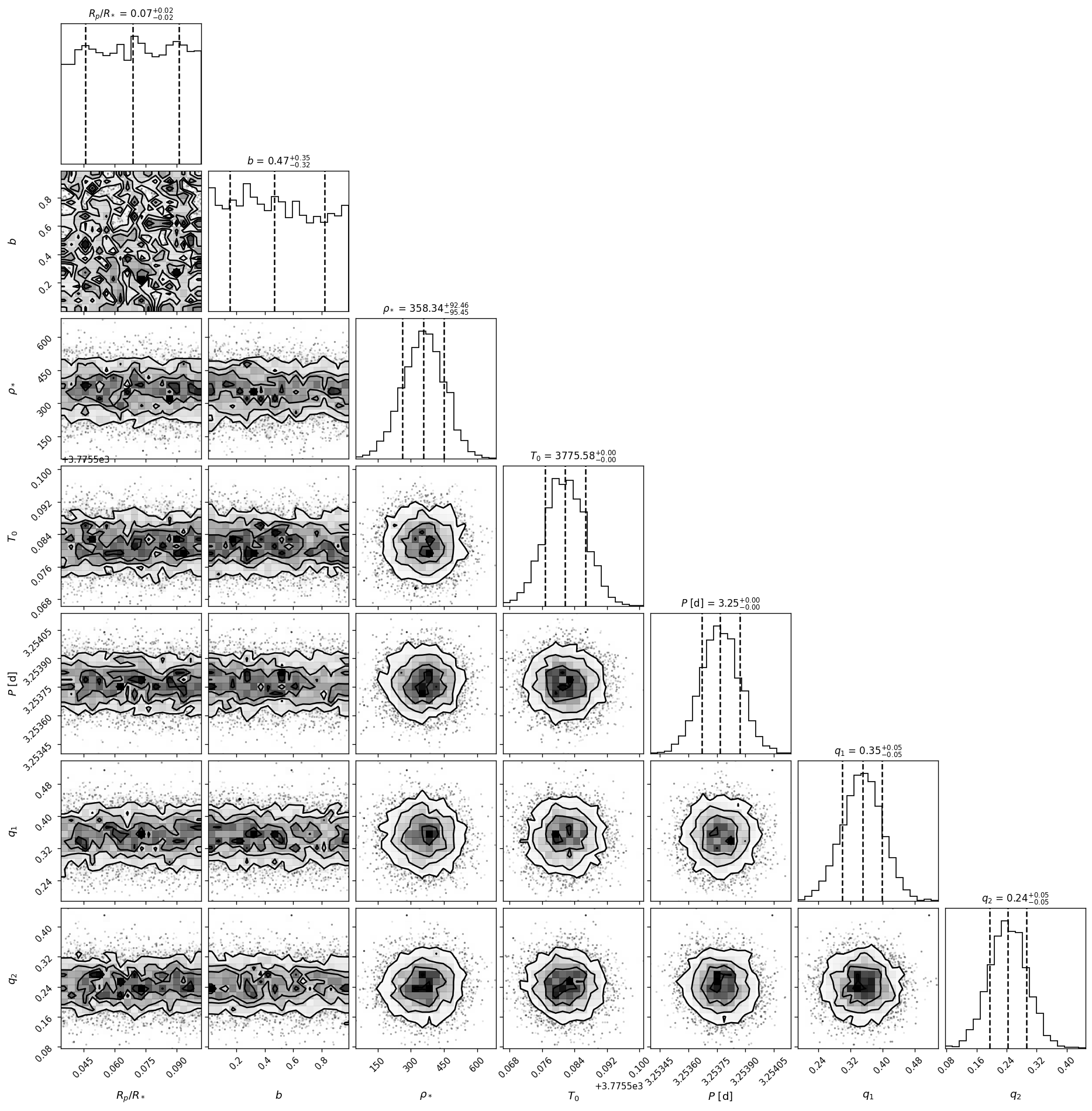}
    \caption{\textbf{Corner Plot of Fit Posteriors.} Diagonal panels: marginalized 1D histograms with the 16th, 50th, and 84th percentiles marked. Off-diagonal panels: 2D joint distributions. Note the broad, essentially flat posterior of $b$ (second column/row) and its correlation with $p = R_p/R_*$, which limits the precision of the radius determination.}
    \label{fig:corner}
\end{figure}

Figure~\ref{fig:corner} shows the corner plot of the fit posteriors. The period and epoch are well-constrained ($\sigma_P \sim 10^{-4}$ days, $\sigma_{T_0} \sim 5 \times 10^{-3}$ days), recovering the BLS solution with minimal improvement, as expected given the tight Gaussian priors.

The radius ratio converges to $p = 0.07^{+0.02}_{-0.02}$. In contrast, the impact parameter $b = 0.46^{+0.36}_{-0.32}$ remains essentially flat across its entire prior range $[0, 0.99]$, indicating that the TESS data cannot distinguish between a central and a grazing transit geometry. This $p$--$b$ degeneracy arises because a larger planet on a grazing orbit produces the same integrated flux decrement as a smaller planet on a central orbit; resolving it requires accurate measurement of the ingress and egress durations, which demands higher cadence or higher photometric precision than available here. As a direct consequence, the stellar density posterior ($358 \pm 93$ kg m$^{-3}$) reproduces the prior almost exactly, since $\rho_*$ is constrained by $a/R_*$ which itself depends on $b$.

\section{Planetary Parameters}
\label{sec:results}

\subsection{Monte Carlo Error Propagation}

Physical parameters were derived via Monte Carlo error propagation with $N = 10{,}000$ samples. At each iteration we drew independently from Gaussian distributions for $M_*$, $R_*$, and $T_\mathrm{eff}$, and from the Juliet posterior samples for $p$, $b$, $\rho_*$, $P$, and $T_0$. This approach correctly propagates the non-Gaussian shape of the posteriors into the derived quantities, avoiding the small-error assumptions of standard analytical propagation. In particular, it accounts for the simultaneous uncertainty in the fit (photometric data) and in the stellar parameters (spectroscopic observations), which would be impossible to combine analytically given their non-linear coupling.

The planet radius follows from $R_p = p \cdot R_*$, converted to Jupiter units using the equatorial ratio $R_\odot / R_\mathrm{Jup} = 9.9549$. The semi-major axis is computed from Equation~(1). The equilibrium temperature assumes Bond albedo $A = 0.3$ and efficient day--night heat redistribution ($f = 0.5$):
\begin{equation}
    T_\mathrm{eq} = T_\mathrm{eff}\left(\frac{R_*}{2a}\right)^{1/2}\!\left[f(1-A)\right]^{1/4}.
\end{equation}

For asymmetric posterior distributions, particularly $M_p$, we report both the median with 16th/84th percentile errors and the Maximum A Posteriori (MAP) value, estimated via Gaussian Kernel Density Estimation. The MAP is more representative of the most probable value when the distribution is strongly skewed.

\subsection{Planet Mass}

In the absence of radial velocity measurements, the planet mass was estimated using the probabilistic mass-radius relation of \citet{chen2017}, implemented via the \texttt{mr\_forecast} package. The resulting posterior is highly asymmetric: the bulk of the probability is concentrated at $M_p \sim 1$--$3\ M_\mathrm{Jup}$, but the distribution extends to $\sim 200\ M_\mathrm{Jup}$ due to the degeneracy in the giant planet regime, where similar radii can correspond to a wide range of masses from standard gas giants to brown dwarfs. The MAP estimate ($M_p \approx 2.2\ M_\mathrm{Jup}$) is therefore more informative than the median ($3.1^{+64.7}_{-2.9}\ M_\mathrm{Jup}$), and should be interpreted as an order-of-magnitude estimate pending RV confirmation.

\subsection{Summary of Parameters}

\begin{figure}[ht!]
    \centering
    \includegraphics[width=1.0\linewidth]{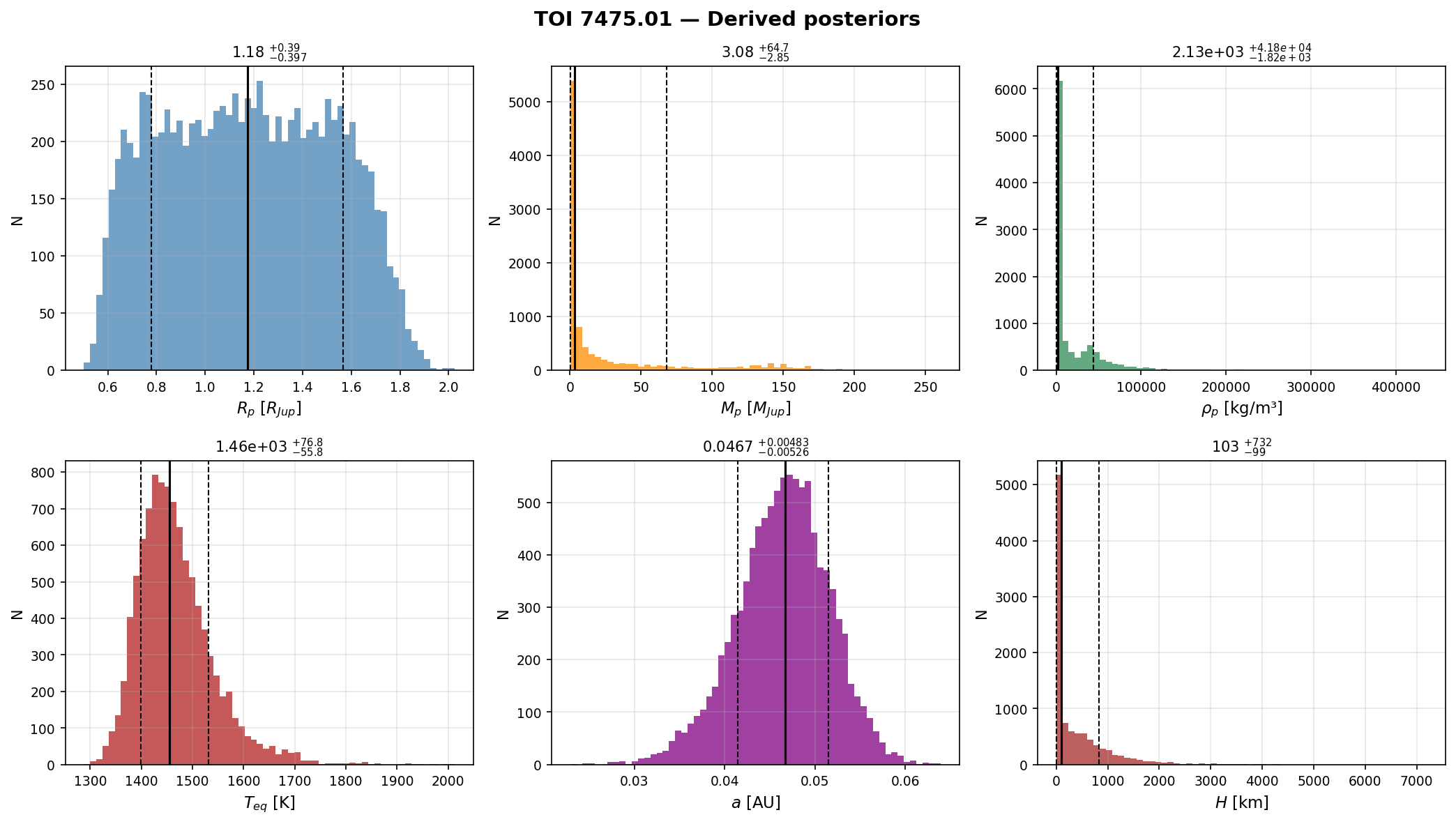}
    \caption{\textbf{Derived Posterior Distributions.} Histograms of the six main physical parameters from the Monte Carlo error propagation ($N = 10{,}000$ samples). Solid vertical lines: median. Dashed lines: 16th and 84th percentiles. The $M_p$ distribution (top center) is strongly asymmetric due to the mass-radius degeneracy in the giant planet regime; the MAP ($\approx 2.2\ M_\mathrm{Jup}$) is more representative than the median.}
    \label{fig:posteriors}
\end{figure}

The complete set of derived parameters is given in Table~\ref{tab:parameters}. The posterior distributions for the main physical quantities are shown in Figure~\ref{fig:posteriors}.

\begin{table*}[ht!]
    \centering
    \caption{Stellar and Planetary Parameters of TOI 7475.01}
    \begin{tabular}{llcc}
        \hline \hline
        Parameter & Symbol & Value & Source \\
        \hline
        \multicolumn{4}{l}{\textit{Stellar parameters}} \\
        Effective temperature & $T_\mathrm{eff}$ [K]     & $6467 \pm 52$       & Spectroscopy \\
        Stellar mass          & $M_*\ [M_\odot]$          & $1.279 \pm 0.200$   & Spectroscopy \\
        Stellar radius        & $R_*\ [R_\odot]$          & $1.726 \pm 0.100$   & Spectroscopy \\
        Surface gravity       & $\log g$ [cgs]            & $4.14 \pm 0.10$     & Spectroscopy \\
        Metallicity           & $[\mathrm{Fe/H}]$         & $-0.10 \pm 0.08$    & Spectroscopy \\
        $J$-band magnitude    & $m_J$                     & $7.676$             & 2MASS \\
        \hline
        \multicolumn{4}{l}{\textit{Orbital parameters (Juliet posteriors)}} \\
        Orbital period        & $P$ [days]      & $3.25377^{+0.00010}_{-0.00009}$ & This work \\
        Transit epoch         & $T_0$ [BJD]     & $2460775.5816^{+0.0050}_{-0.0050}$ & This work \\
        Eccentricity          & $e$             & $0$ (fixed)              & This work \\
        Radius ratio          & $p = R_p/R_*$   & $0.07^{+0.02}_{-0.02}$   & This work \\
        Impact parameter      & $b$             & $0.46^{+0.36}_{-0.32}$   & This work \\
        \hline
        \multicolumn{4}{l}{\textit{Planetary parameters (Monte Carlo)}} \\
        Planet radius (median)    & $R_p\ [R_\mathrm{Jup}]$      & $1.18^{+0.39}_{-0.40}$  & This work \\
        Planet radius (MAP)       & $R_p\ [R_\mathrm{Jup}]$      & $1.18$                  & This work \\
        Planet mass$^\dagger$ (MAP) & $M_p\ [M_\mathrm{Jup}]$    & $\approx 2.2$           & \citet{chen2017} \\
        Semi-major axis           & $a$ [AU]                      & $0.047^{+0.005}_{-0.005}$ & This work \\
        Equilibrium temperature   & $T_\mathrm{eq}$ [K]           & $1455^{+77}_{-56}$      & This work \\
        Incident flux             & $F_p\ [F_\oplus]$             & $2137^{+488}_{-309}$    & This work \\
        Planet density (MAP)      & $\rho_p$ [kg m$^{-3}$]        & $\approx 1482$          & This work \\
        Atm. scale height (MAP)   & $H$ [km]                      & $\approx 29$            & This work \\
        TSM ($M_p = 1\ M_\mathrm{Jup}$) & ---                    & 17.1                    & \citet{kempton2018} \\
        \hline
    \end{tabular}
    \tablecomments{$^\dagger$ Mass estimated from the \citet{chen2017} probabilistic M-R relation. The MAP value is reported as it is more representative than the median for this highly asymmetric distribution. Radial velocity measurements are required for a precise determination.}
    \label{tab:parameters}
\end{table*}

\section{Discussion}
\label{sec:discussion}

\subsection{Classification}

Both the median and MAP radius estimates agree ($R_p \approx 1.18\ R_\mathrm{Jup}$), placing the planet in the Jupiter-sized category. The 1$\sigma$ range spans $0.78$--$1.57\ R_\mathrm{Jup}$, entirely within the giant planet regime, so the size classification is robust despite the large formal uncertainty. The equilibrium temperature $T_\mathrm{eq} = 1455^{+77}_{-56}$ K is well-constrained, as it depends primarily on $a$ and $T_\mathrm{eff}$, both of which are well-determined, and firmly places the system in the hot Jupiter regime ($T_\mathrm{eq} > 1000$ K).

The MAP density ($\rho_p \approx 1482$ kg m$^{-3}$) is consistent with a non-inflated hot Jupiter ($\rho_\mathrm{Jup} = 1326$ kg m$^{-3}$). This is noteworthy given the high irradiation ($F_p \approx 2137\ F_\oplus$), as ohmic heating and other inflation mechanisms are expected to be active at this flux level \citep{demory2011}. However, the mass uncertainty is sufficiently large that this conclusion is not firm: a lower actual mass would imply a lower density and a more inflated structure.

\subsection{Population Context}

\begin{figure}[ht!]
    \centering
    \includegraphics[width=1.0\linewidth]{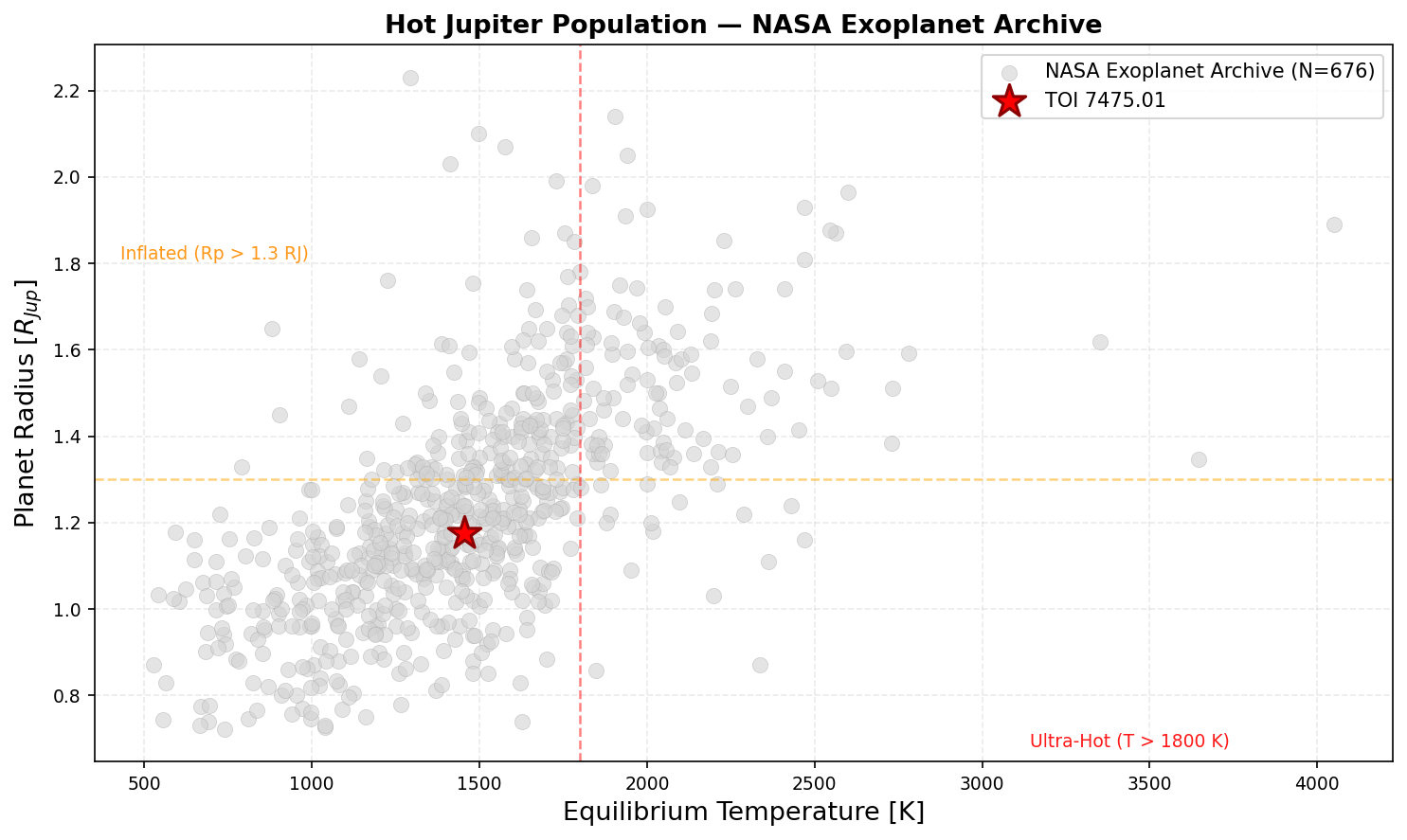}
    \caption{\textbf{Hot Jupiter Population Comparison.} Gray points: 676 hot Jupiters from the NASA Exoplanet Archive with known radii and equilibrium temperatures. Red star: TOI 7475.01 (median values). Dashed lines mark the inflated ($R_p > 1.3\ R_\mathrm{Jup}$) and ultra-hot ($T_\mathrm{eq} > 1800$ K) boundaries. TOI 7475.01 sits in a well-populated region at intermediate temperature and typical Jupiter size.}
    \label{fig:population}
\end{figure}

Figure~\ref{fig:population} places TOI 7475.01 in the context of 676 hot Jupiters ($P < 10$ days, $R_p > 0.5\ R_\mathrm{Jup}$, $T_\mathrm{eq} > 500$ K) from the NASA Exoplanet Archive. The planet occupies a well-populated region at intermediate temperature ($\sim 1450$ K) and typical Jupiter radius, broadly consistent with the bulk of the known hot Jupiter population.

\subsection{Impact Parameter Degeneracy and Recommended Follow-up}

The principal limitation of this characterization is the unconstrained impact parameter ($b = 0.46^{+0.36}_{-0.32}$), which propagates into a large fractional radius uncertainty ($\sim 34\%$). This degeneracy is a known limitation of single-band TESS photometry and arises because the integrated transit depth is preserved across a family of $(p,b)$ solutions with the same $p^2$. As visible in the corner plot (Figure~\ref{fig:corner}), the posterior of $b$ is essentially flat over its entire prior range, and the joint $p$--$b$ distribution shows no constraining power from the data alone.

High-precision follow-up photometry with CHEOPS \citep{benz2021} would break this degeneracy by resolving the ingress and egress shapes with sufficient precision to separately constrain $b$ and $p$. Ground-based multi-band photometry could additionally constrain the limb darkening independently, providing a further check on the stellar parameters.

Radial velocity observations are equally critical. The 1$\sigma$ mass range from the \citet{chen2017} relation spans $0.23$--$68\ M_\mathrm{Jup}$, making it impossible to determine whether the companion is a planet, a massive giant, or a brown dwarf without direct Doppler measurement.

\subsection{Atmospheric Observability}

The Transmission Spectroscopy Metric \citep[TSM,][]{kempton2018} for this system is TSM $\approx 17$ (assuming $M_p = 1\ M_\mathrm{Jup}$), below the recommended threshold of 90 for hot Jupiter atmospheric characterization with JWST. The low TSM reflects the brightness of the host star in the near-infrared ($m_J = 7.676$) combined with the moderate planet-to-star radius ratio. While the atmospheric scale height $H \approx 29$ km (MAP) is typical for hot Jupiters, this planet is a marginal JWST target compared to the best-characterized systems currently known.

\section{Conclusion}
\label{sec:conclusion}

Based on the high SNR detection (294.13), the clean spatial environment, stable centroid analysis, and a False Positive Probability of FPP $\approx 0$, we confirm that \textbf{TOI 7475.01 is a statistically validated exoplanet}. The signal exhibits all the characteristics of a genuine planetary transit, with no evidence of contamination from background sources or binary stars.

The subsequent Bayesian characterization yields the following main results:

\begin{enumerate}
    \item Planet radius $R_p = 1.18^{+0.39}_{-0.40}\ R_\mathrm{Jup}$, consistent with a Jupiter-sized body across the full 1$\sigma$ range.
    \item Equilibrium temperature $T_\mathrm{eq} = 1455^{+77}_{-56}$ K, firmly classifying the system as a hot Jupiter.
    \item Estimated mass $M_p \approx 2.2\ M_\mathrm{Jup}$ (MAP, Chen-Kipping), pending RV confirmation.
    \item The impact parameter $b$ is unconstrained by the TESS data alone, limiting the precision of the radius determination. This is the principal observational limitation of this work.
    \item CHEOPS follow-up photometry is recommended to break the $p$--$b$ degeneracy, and radial velocity observations are required to measure the true planetary mass.
\end{enumerate}

\section*{Data Availability}

The TESS data presented in this paper can be obtained from the Mikulski Archive for Space Telescopes (MAST) at the Space Telescope Science Institute (STScI). The specific data products used (Sector 91) are publicly available under the target TIC 376866659. The analysis pipeline is publicly available at \url{https://github.com/biesro/TESS-TOI-7475.01-Validation}.

\section*{Acknowledgments}

This paper includes data collected by the TESS mission, which are publicly available from the Mikulski Archive for Space Telescopes (MAST). We acknowledge the use of public data from the \textit{Exoplanet Follow-up Observation Program} (ExoFOP), which is operated by the NASA Exoplanet Science Institute.

This research made use of \texttt{Lightkurve} \citep{lightkurve2018}; \texttt{TRICERATOPS} \citep{giacalone2021}; \texttt{juliet} \citep{espinoza2019}; Dynesty \citep{speagle2020}; \texttt{mr\_forecast} \citep{chen2017}; \texttt{corner.py} \citep{corner2016}; \texttt{Astropy} \citep{astropy2022}; \texttt{NumPy} \citep{harris2020}; \texttt{Matplotlib} \citep{hunter2007}; and \texttt{pandas}.


\end{document}